# Neuromelanin-MRI using 2D GRE and deep learning: considerations for improving the visualization of substantia nigra and locus coeruleus

*Samy Abo Seada[1], Anke W. van der Eerden[1], Agnita J.W. Boon[2], Juan A. Hernandez-Tamames[1,3]*

[1] *Department of Radiology and Nuclear Medicine, Erasmus MC, Rotterdam, The Netherlands*
[2] *Department of Neurology, Erasmus MC, Rotterdam, The Netherlands*
[3] *Department of Imaging Physics, TU Delft, The Netherlands*

Corresponding author:
Juan Hernandez Tamames
Email: j.hernandeztamames@erasmusmc.nl

## Abstract

An optimized clinically feasible neuromelanin-MRI imaging protocol for visualising the SN and LC simultaneously using deep learning reconstruction is presented. We optimize flip-angle for optimal combined SN and LC depiction. We also experimented with combinations of anisotropic and isotropic in-plane resolution, partial vs full echoes and the number of averages. Phantom and in-vivo experiments on three healthy volunteers illustrate that high-resolution imaging combined with deep-learning denoising shows good depiction of the SN and LC with a clinically feasible sequence of around 7 minutes.

## Introduction

Neuromelanin is a dark pigment primarily found in the Substantia Nigra (SN) and Locus Coeruleus (LC). The pigment consists of organic compounds as well as metal ions, such as iron[1]. Imaging neuromelanin with MRI is possible in number of ways, commonly with either magnetization transfer (MT)-weighed gradient echo or $T_1$-Fast Spin Echo (FSE) sequences. Several studies have shown its usefulness in the diagnosis of idiopathic Parkinson's Disease [2-5] and detecting early-stage Parkinson's disease[6, 7]. Few studies showed some usefulness for diagnosing atypical parkinsonism[3, 8] but further clinical use remains a research question[9-11]. Neuro-melanin concentration was also found to positively correlate with the severity of psychosis in patients with schizophrenia[12]. An advantage for this imaging technique is that it requires no further post-processing, however standardizing volume or contrast-ratio (CR) measurements is still a barrier to clinical translation.

The LC is a rod-shaped structure only a few mm in diameter, running parallel to the back of the brainstem. The substantia nigra (SN) is a basal ganglia structure which lies in the midbrain, where neuromelanin lies in dopaminergic neurons, mostly within a substructure known as SN pars compacta (SNpc). Several studies have used MRI-based metrics which rely on contrast ratios[3, 13, 14], area[15] or volume [2, 6] for quantifying neuromelanin in the SN and/or LC between different patient or healthy populations. The common ground for these metrics is that they are intensity-based metrics, which rely on good delineation between the signal of interest (neuromelanin-rich tissue) and the background signal, which is the surrounding midbrain and pons tissue.

Acquisition with MRI is performed by placing the imaging window axially obliqued such that the plane is perpendicular to the dorsal side of the brainstem. These images are inherently low in SNR, and are conventionally acquired multiple times and averaged, to increase image SNR. Increasing averages directly increases acquisition time, which has been reduced by using acceleration techniques such as parallel imaging, partial echo[16] (Partial fourier in the readout direction) and by reducing the required number of phase encoding steps, which leads to anisotropic inplane resolution. We reviewed several acquisitions from published NM-MRI articles, and found a range of strategies for achieving optimal contrast or increasing reproducibility. Using parallel imaging is uncommon, and the popularity of anisotropic acquisition resolution is uncertain, since many studies report voxel size rather than matrix size. However, it has been used explicitly in several reports[2, 3, 16].

In this work we investigated the effects from some of the acquisition parameters on the visualization of neuromelanin in the SNpc and LC. Specifically, we investigated varying flip-angle and three SNR-related parameters, namely partial echo, image resolution and the number of averages, and focused on how these can be optimized for visualising neuromelanin in the SNpc and the LC. A k-space illustration of the different parameters is shown in Figure 1. We also investigated the use of vendor-provided deep-learning based denoising techniques and demonstrate results using phantom and in-vivo experiments.

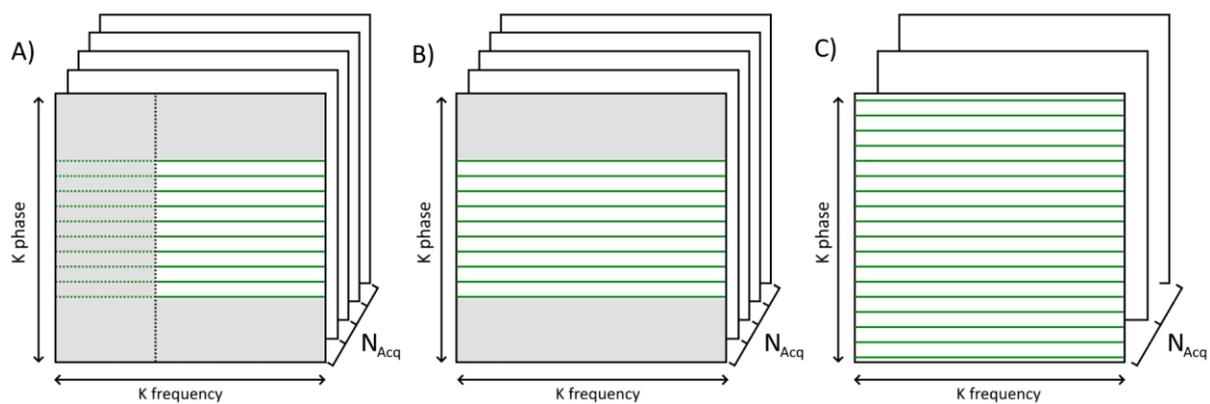

*Figure 1: A k-space illustration of the comparison made in this work. A) shows an illustration of a typical 2D GRE for Neuromelanin-MRI can be acquired quickly. The shaded k-space regions are not acquired which leads to lower resolution in the phase direction. The stippled green lines are the segments of k-space not acquired in the readout direction when partial echo is enabled. To improve SNR, the acquisition is repeated several times (Nacq) and averaged. Subfigure B) shows the k-space when partial echo is disabled. Subfigure C) shows the k-space when high-resolution is acquired in both the frequency and phase dimension, as well as fully acquired echos. This will almost certainly improve image quality, so to make a fair comparison the number of averages is reduced to approximately match acquisition time.*

## Theory and Methods

This section covers the theory and introduces our experiment design. Our experiments aim to answer three questions on optimising neuromelanin MRI for the sake of visualising its content in the SN and LC.

1) NM-MRI contrast mechanisms and setting the optimal flip-angle optimization for in-vivo experiments
2) Balancing SNR and isotropic resolution, including
    a. Acquiring with partial vs full echoes
    b. Acquiring with anisotropic vs isotropic in-plane resolution
    c. Adjusting number of acquisitions
3) Investigating the benefits of deep-learning based resolving and denoising reconstruction

## Contrast mechanisms and flip-angle optimisation

The MRI contrast mechanisms which govern the contrast between NM-rich tissue and the surrounding pons are believed to be a combination of (MT) and $T_1$-shortening effect due to the iron present in NM-pigments[1]. The MT effect is a consequence of RF radiation in the bound pool, which saturates the MRI signal from protons in the free pool, which is dominated by water. RF radiation in the bound pool is achieved by transmitting RF pulses either at on and off-resonance frequencies, or both, where off-resonance frequencies typically vary between 600-2000Hz for 3T.

$T_1$-shortening is a term describing the signal enhancement of certain tissues in $T_1$-weighted imaging due to the presence of paramagnetic components. Most commonly this is used in the context of gadolinium and its effect on $T_1$-weighted imaging, however the presence of an iron-element in the neuromelanine complex makes the term applicable to neuromelanine-MRI as well.

Both TSE and GRE sequence families can be set up for having high $B_1$-RMS (for MT) and short echo-times (for $T_1$-weighting) to satisfy these conditions. In this study, we used an MT-weighted 2D GRE sequence which was found to be produce improved results over 3D techniques during initial results, as well as in used in other studies[14, 16, 17].

Changing the flip-angle can dramatically affect neuromelanin contrast in relation to its surrounding tissue, especially for 2D GRE acquisition where the minimum sequence TR varies with the number of selected slices. A previous study [18] at 3T performed relaxometry and used the SPGR-signal equation to optimise the flip-angle for a 3D acquisition. Here we use the same relaxometry values for a 2D acquisition and compare our calculations with in-vivo data. We acquired the in-vivo data with isotropic resolution and full echoes, to avoid conflating imaging artifacts.

## Balancing SNR and isotropic resolution

Image SNR, image resolution and acquisition time are interdependent parameters which need to be considered together when optimising a sequence. Partial echo can be used to accelerate acquisition, by acquiring a segment of k-space in the readout direction and approximating the non-acquired segment by relying on the conjugate symmetry of k-space[19]. Partial echoes allow for shorter TE and a shorter TR, however reduces the SNR, and relies on the symmetry of echoes which holds when the $T_2$ values are much longer than the echo readout duration.

Starting from an isotropic in-plane acquisition resolution, moving to an anisotropic resolution in the phase-encoding direction reduces scan-time, but can results partial volume effect in the same direction, which could hamper depicting small structures. Note that the acquisition resolution is not always identical to the reconstruction resolution, as often MRI images are interpolated to a higher resolution during reconstruction. The number of phase encoding steps can also be reduced using parallel imaging at a cost in SNR. However, this is not commonly used in NM-MRI protocols, and instead multiple acquisition are used to increase SNR.

In one set of phantom experiments we used partial and full echoes, as well as isotropic and anisotropic resolution, for a single acquisition case, to separate imaging effects from SNR increases. In a second set of experiments, we used $N_{acq}$=3 for the acquisitions with isotropic resolution and $N_{acq}$=5 for the one with anisotropic resolution, to match in-vivo acquisition times (within 4%), and repeated this experiment in-vivo.

### Deep-learning based reconstruction

Another approach to increase image quality is by using machine-learning based resolving and denoising features, introduced recently by major MRI vendors[20]. We explored how this would affect image quality on our GE scanner.

### Experiments

Data was acquired on a 3T GE Signa Premier (GE Healthcare, Chicago, USA) using a 48 channel head-coil. Three healthy volunteers were recruited according to local ethics after informed consent was acquired. For the flip-angle optimisation experiment, we used a 2D MT-weighted GRE sequence with a resolution of 0.4x0.4x3mm$^3$, 512x512, 12 slices, TE=7.5ms, TR=340ms, vendor MT-pulse 8ms duration, 1200Hz offset, and no intensity correction. The flip-angle was varied from 30º to 70º in steps of 10º.

For the SNR vs isotropic resolution experiment, a system phantom (CaliberMRI, Boulder, Colorado, USA) was scanned coronally at the height of a resolution inset plate with structures ranging from 0.8 to 0.4mm. The first set ($N_{acq}$=1) of experiments was acquired using an identical sequence with FA=50º, using a single slice and a manually set TR of 340ms to avoid variations in contrast, apart for the relatively small change of TE. Partial echo was enabled which reduced the TE to 4.8ms, and anisotropic resolution (0.4x0.7mm$^2$) was enabled by setting the acquisition matrix to 512x320. Frequency encoding was in AP, and intensity correction was enabled. In the second set ($N_{acq}$>1) of experiments, $N_{acq}$=5 was set for the anisotropic acquisitions and $N_{acq}$=3 for the isotropic resolution, to match the in-vivo comparison to follow. The same 4 sequences used for phantom were used for the in-vivo acquisitions, however with 12 slices and minimum TE enabled, which led to the following acquisition times:

1. 0.4x0.7mm$^2$; 512x320; partial echo; Nacq=5; TE:4.0ms; TR:300ms; Acq time 6:52s
2. 0.4x0.7mm$^2$; 512x320; full echo;    Nacq=5; TE=7.5ms; TR=340ms; Acq time 7:46s
3. 0.4x0.4mm$^2$; 512x512; partial echo; Nacq=3; TE:4.0ms; TR:300ms; Acq time 6:36s
4. 0.4x0.4mm$^2$; 512x512; full echo;    Nacq=3; TE=7.5ms; TR=340ms; Acq time 7:28s

All sequences were acquired with the vendor-based DL Recon set to high and the reconstruction matrices set to 1024x1024, and the original images prior to DL recon were saved.

### Results

Figures 2 show simulated Ernst-signal curves for two different TRs with red dots highlighting the optimal operating points. The change in optimal flip-angle as a function of TR highlights the necessity to optimize based on the number of acquired slices.

Figure 3 shows in-vivo demonstration from flip-angle optimisation. We used the numerical calculation from 12 slices (TR=340ms) and varied the sequence flip-angle. For NM in the SN, we found that the optimal flip-angle would be 30-40 degrees (A, red dot) which seems to correspond with in-vivo results (D). For the LC, the ideal flip-angle as determined by the difference between the LC and gray matter, would be 30 degrees (B, red dot). However, due to the CSF signal being high in this regime, it renders the LC less visible. It is therefore better to acquire at a higher flip-angle. In compromise, a flip angle around 50 seems preferable for seeing both structures.

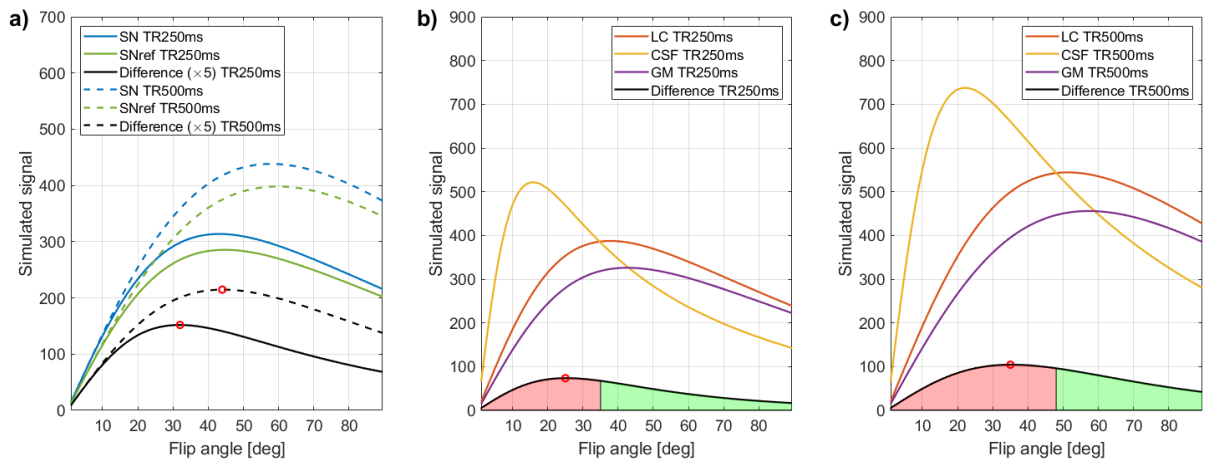

*Figure 2:* a) Shows signal intensity calculations for a TR of 250ms (solid lines) and 500ms (dashed lines), and their difference curves in black. The optimal flip angle for maximizing contrast is marked by a red circle, and is achieved at a flip angle of 32 and 43 for a TR of 250ms and 500ms respectively. For a TR in the middle, one can also expect an optimal Flip angle inbetween this range. For LC, determing the optimal flip-angle is more complex. Considering only the neighbouring gray matter around the LC, one would need a low flip-angle. However, the neighbouring CSF in the 4th ventricle produces a large amount of signal, making the LC less visible, which is shown by the red patch. Choosing a flip-angle above where the CSF and LC curve intersect (green patch) will avoid CSF signal contamination.

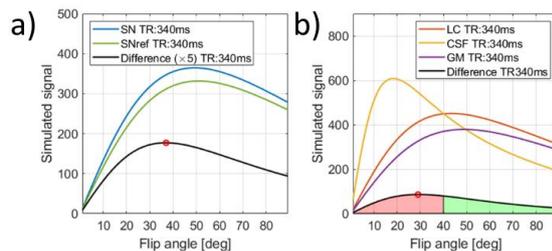

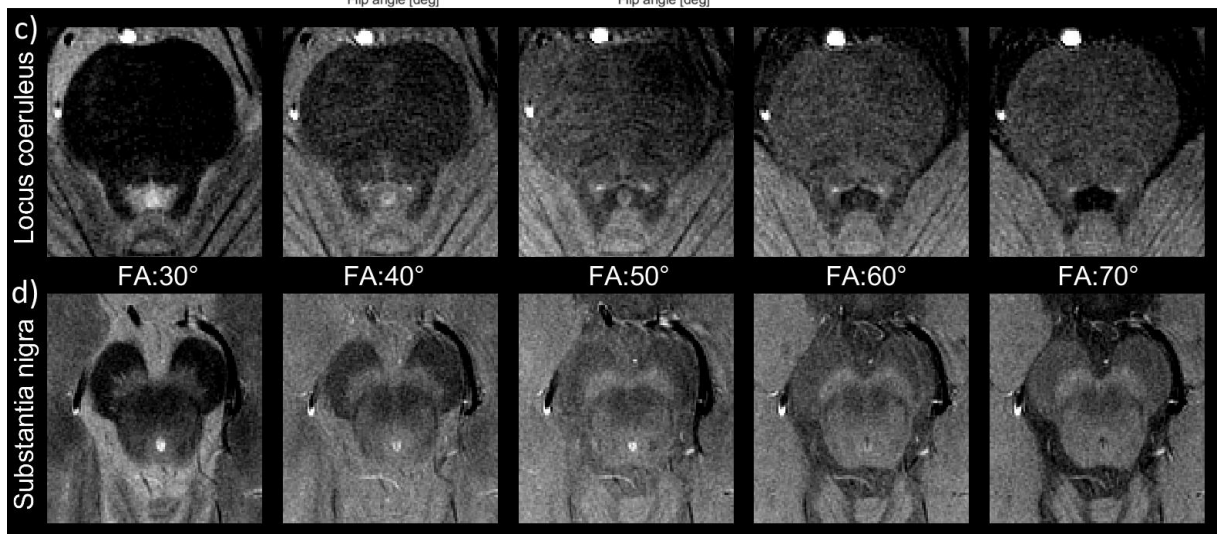

**Figure 3**: *In-vivo demonstration from flip-angle optimisation. We used the numerical calculation from 12 slices (TR=340ms) and varied the sequence flip-angle. For NM in the SN, we found that the optimal flip-angle would be 30-40 degrees (A, red dot) which seems to correspond with in-vivo results (D). For the LC, the ideal flip-angle as determined by the difference between the LC and gray matter, would be 30 degrees (B, red dot). However, due to the CSF signal being high in this regime, it renders the LC*

*less visible. It is therefore better to acquire at a higher flip-angle. In compromise, a flip angle around 50 seems preferable for seeing both structures.*

Figure 4 shows the results from a single-average vs multiple-average experiment, and figures 5-9 show in-vivo results from two healthy volunteers. See the detailed figure legends and discussion for an analysis.

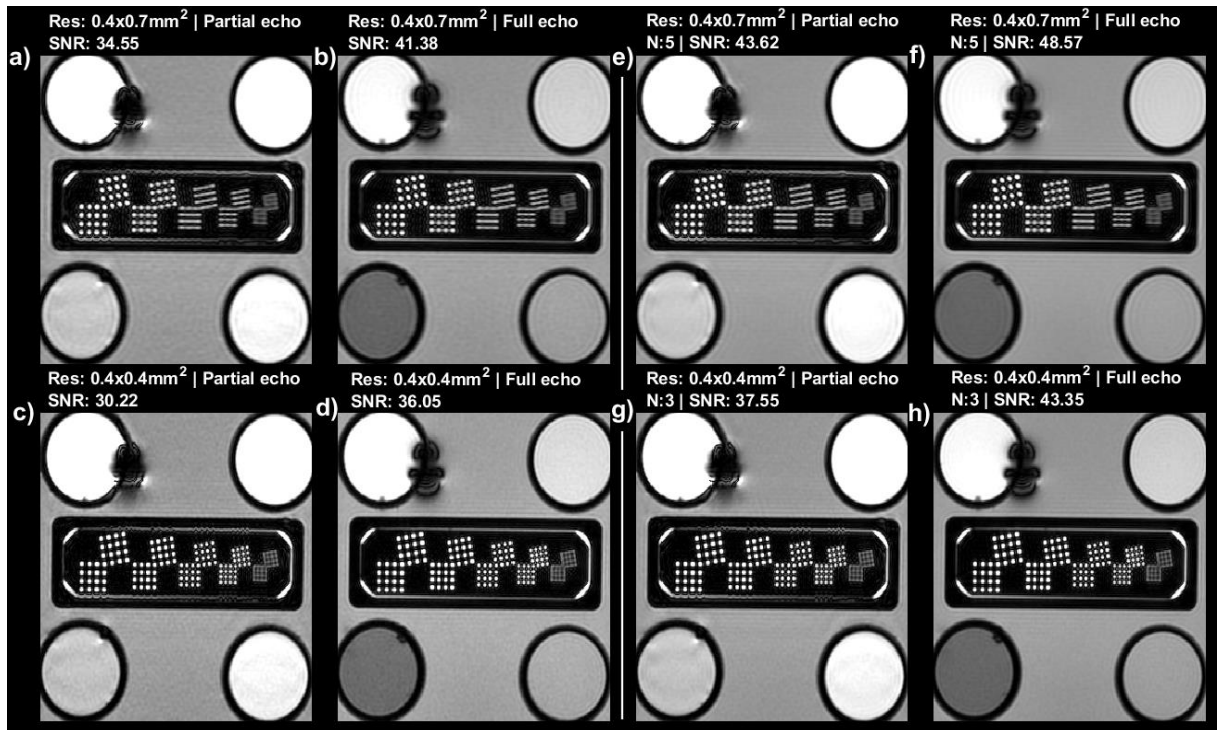

**Figure 4**: *Phantom results from SNR vs isotropic resolution experiments. Subfigures a-d) show the results for the single-average experiment ($N_{acq}=1$). Disabling partial echo (a->b and c->d increases SNR and reduce susceptibility artifacts, at the cost of acquisition time. Moving to isotropic resolution increases acquisition time and reduces SNR, however, makes small structures more visible. This can be seen by the dots that seem connected in a and b, becoming delineated in c and d along the phase-encoding direction. The SNR benefit from (a->d) seems like an improvement, but is a not fair since d has the longest acquisition per slice. Subfigures e-h) show similar results with multiple averages ($N_{acq}>1$), where the N=5 and N=3 acquisitions took 9:07s and 8:46s with a set TR of 340ms. Subfigures e and f show that increasing acquisitions does not delineate the structures either, highlighting that depiction depends on high resolution and not on SNR. Moving towards isotropic resolution (f->h) reduces SNR, and disabling partial echo (g->h) delivers an image with clear depiction and free of artifacts. When accounting for multiple averages, the SNR estimate from the initial sequence is almost unchanged (e->h).*

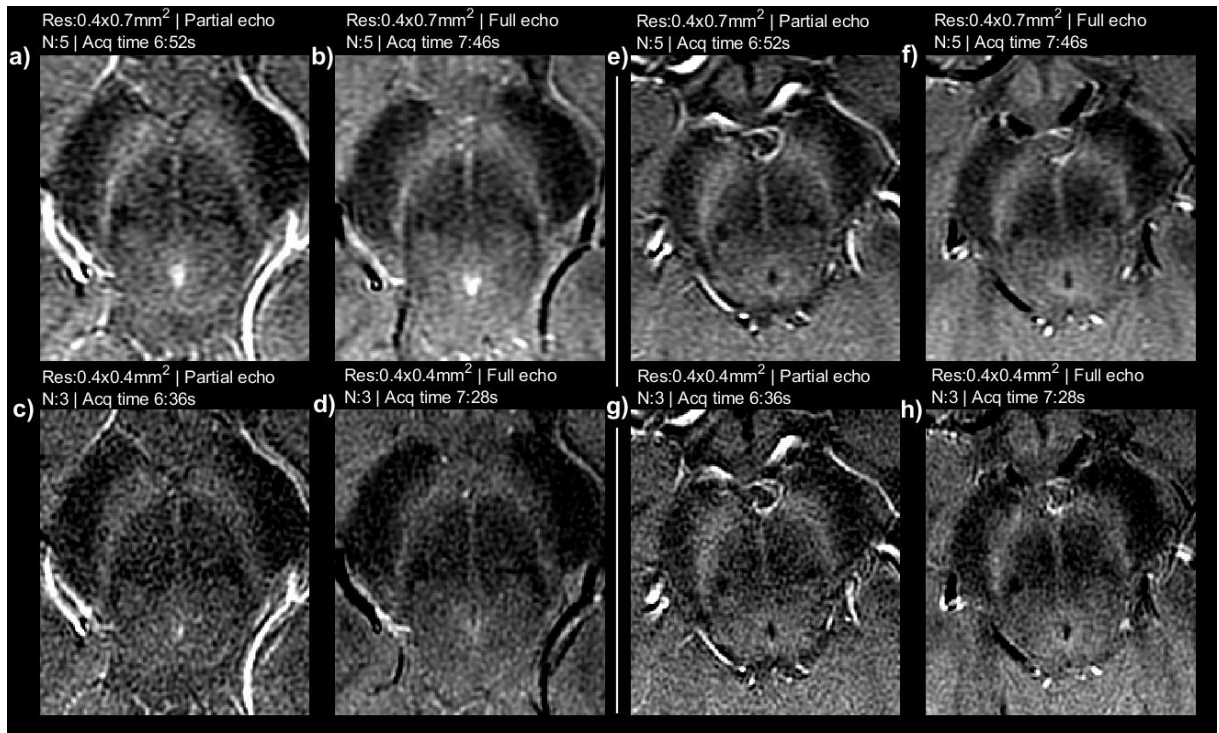

**Figure 5**: *In-vivo comparison for SNR vs isotropic resolution, for subject 1 (a-d) and 2 (e-h) at the substantia nigra. The top row shows low-resolution images with 5 averages, and the bottom row shows high-resolution images with 3 averages, with roughly similar acquisition time (within 4%). Moving from partial to full echoes removes susceptibility artifacts from neighbouring vessels and increases SNR, which is most clearly seen from the reduction of noise in hypointense tissue neighbouring the SNpc. Higher-resolution images show more detailed structures, but are clearly noisier than their lower-resolution counterparts.*

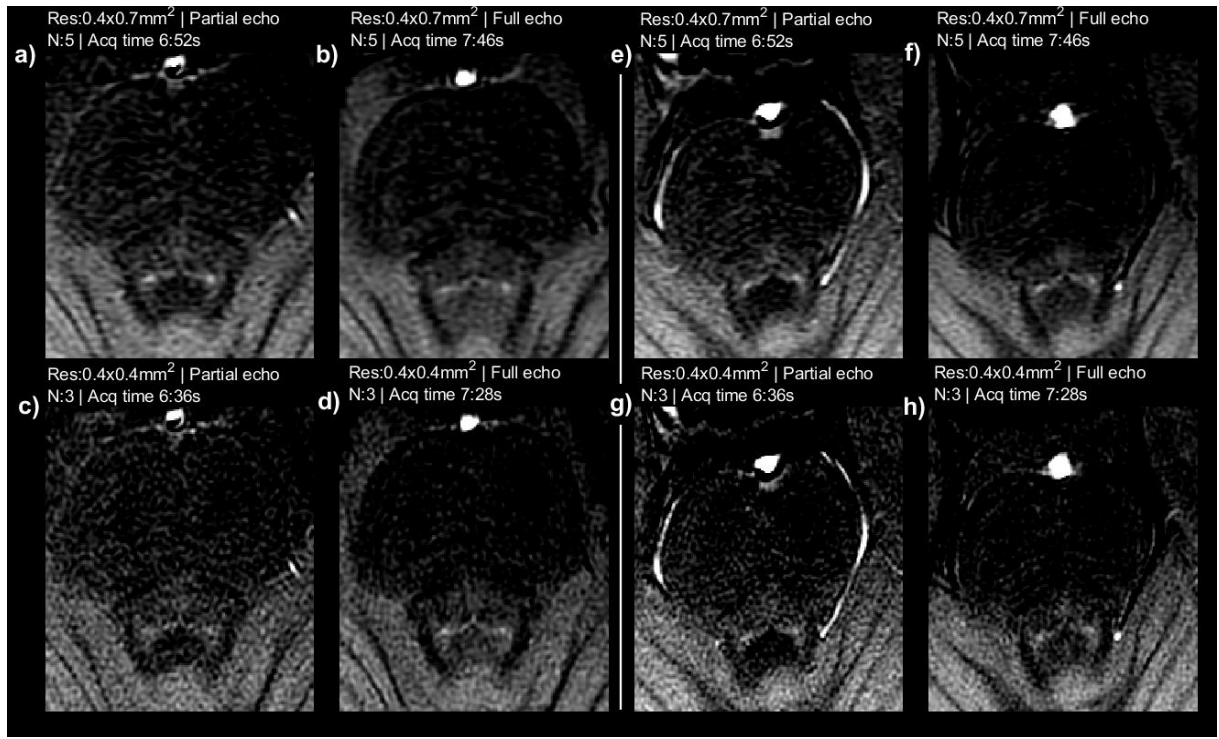

**Figure 6:** *In-vivo comparison for SNR vs isotropic resolution, for subject 1 (a-d) and 2 (e-h) at the locus coeruleus. The top row shows low-resolution images with 5 averages, and the bottom row shows high-resolution images with 3 averages, with roughly similar acquisition time (within 4%). Moving towards isotropic resolution again reduces SNR, however makes the locus coeruleus appear more clearly as a dot.*

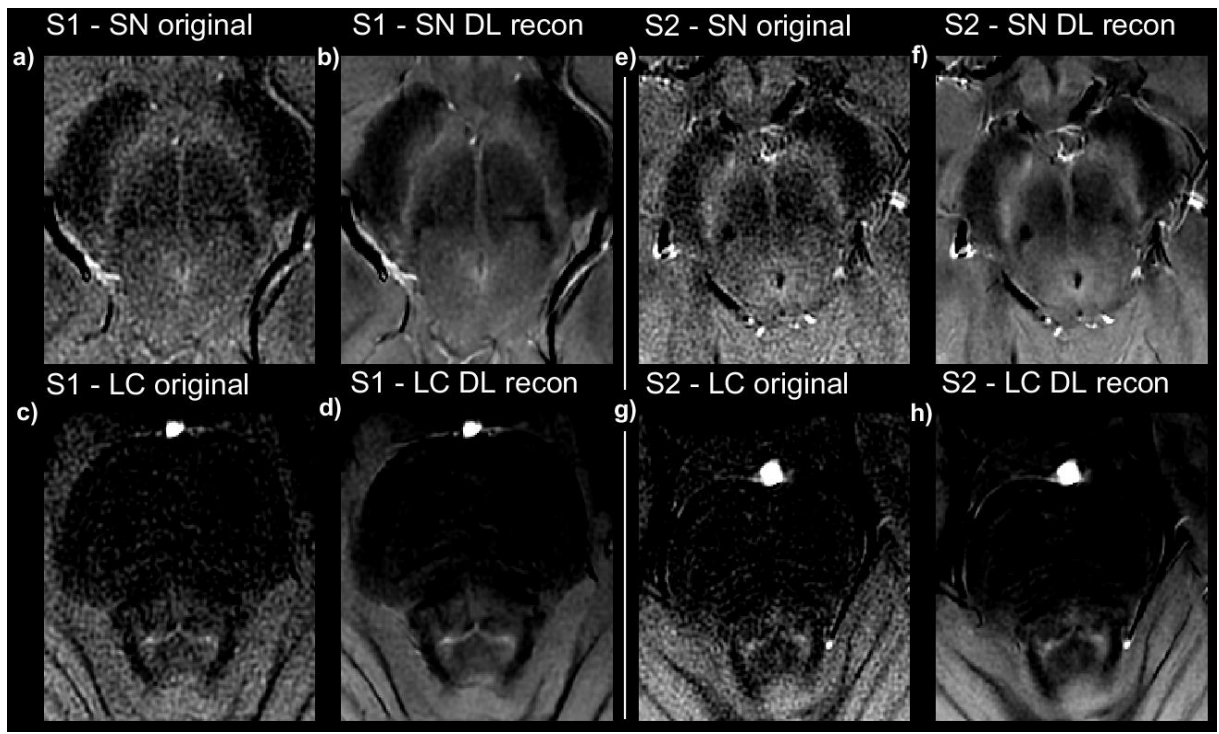

**Figure 7**: *Original (0.4x0.4mm², full-echo) and DL-recon images at the substantia nigra and locus coeruleus for subject 1 (a-d) and 2 (c-d). Images are clearly denoised after deep-learning denoising, however depiction of anatomical structures seem unchanged. This might be of good benefit for CNR-based methods which use signal variance in reference regions.*

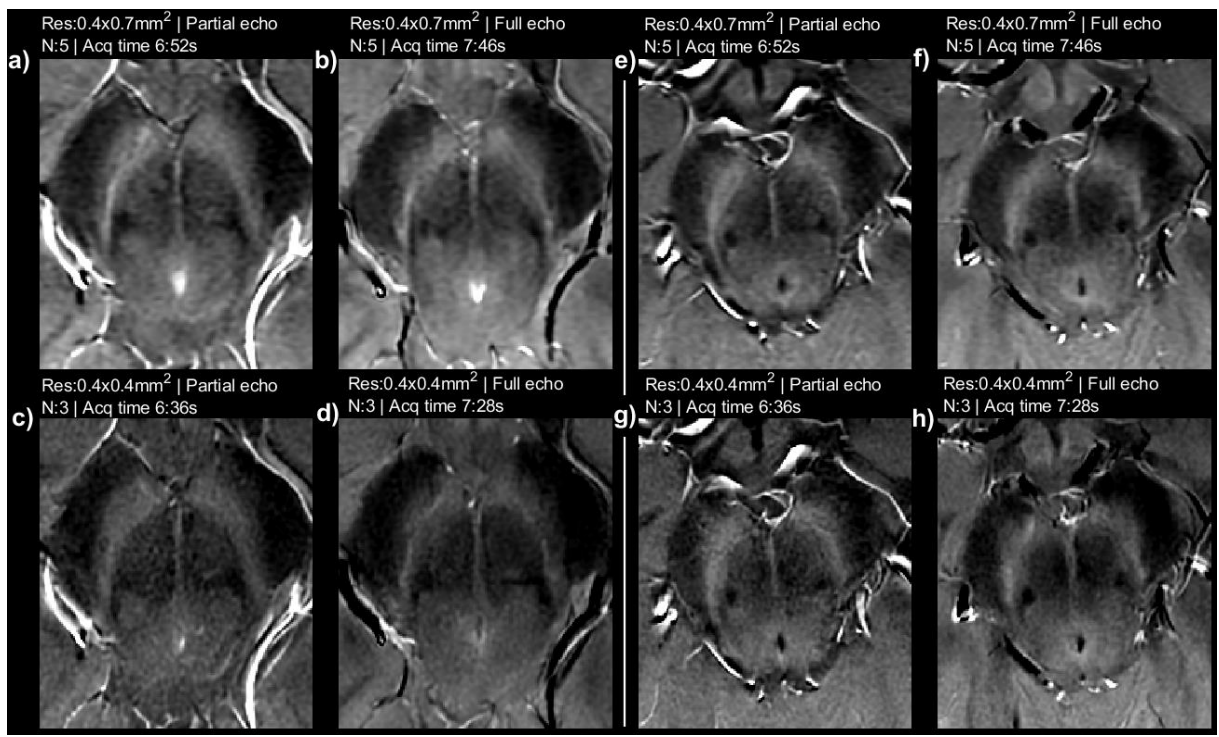

**Figure 8**: *All DL-recon images at the substantia nigra and locus coeruleus for subject 1*

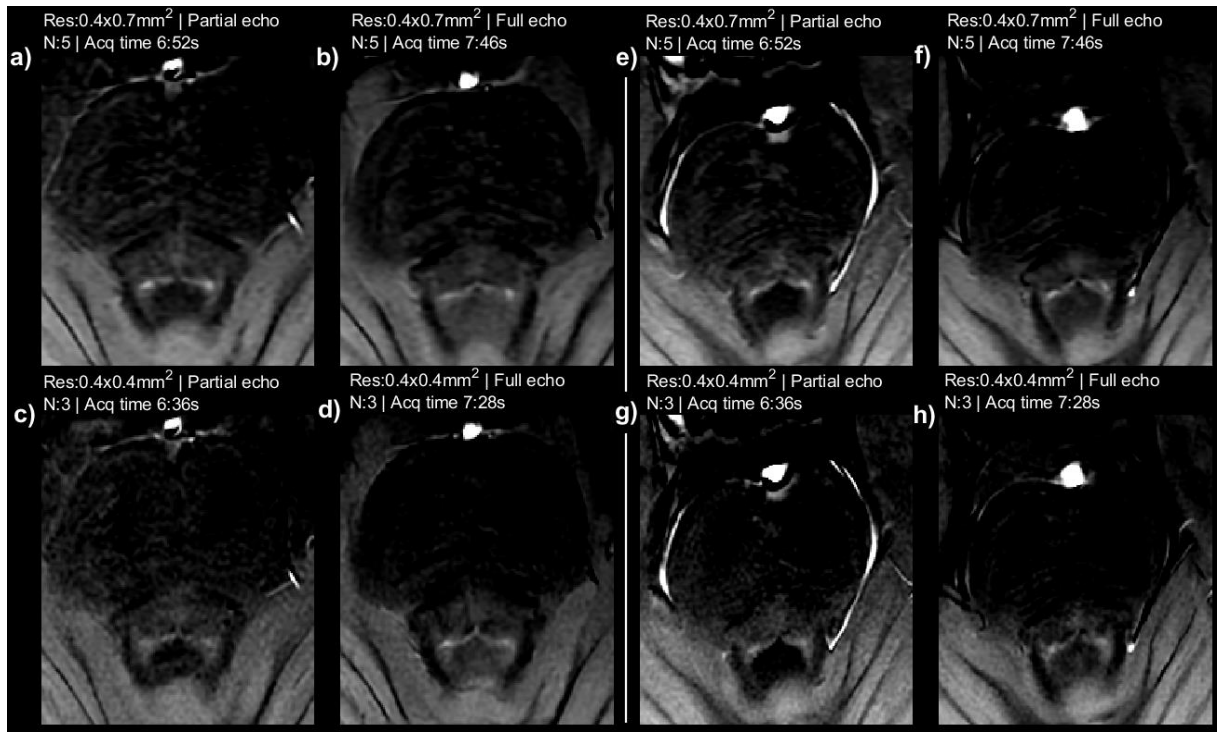

**Figure 9**: *All DL-recon images at the substantia nigra and locus coeruleus for subject 2*

## Discussion

Figures 2 and 3 show results from our flip-angle optimization for improving visualising neuromelanin in the SN and LC simultaneously. We found that for optimal visualization it is important to set the flip-angle based on TR, and hence the number of slices in a 2D GRE sequence. Furthermore, based on 3T relaxometry values and a single in-vivo dataset, we found that there is a compromise to be made between visualizing the SN and LC simultaneously.

Figure 4 shows phantom results from a resolution insert acquired at a range of parameters which affect resolution and SNR. Acquiring full echoes gives an SNR benefit and reduces susceptibility artifacts, as expected. In addition, acquiring with an isotropic in-plane acquisition improves visibility of small dots at a cost in acquisition time and SNR. However, when accounting for multiple averages (Figure 4 e-h) it was found that the combination of enabling full echoes, acquiring with fewer averages and increased resolution results in a similar SNR as partial echoes with more averages and anisotropic resolution. The retained benefit with the latter approach is fewer susceptibility artifacts and better delineation (Figure 4e vs Figure 4h).

Figures 5 to 9 show repetitions of the same principle on healthy subjects at the SN and LC respectively. Using partial echoes increases susceptibility artifacts, seen by hyperintense signal from vessels. The SN is more visible with partial echo and lower resolution (Figure 5a,b) however, finer substructures can be seen at higher resolution, albeit with worse resolution. Adding vendor-based deep-learning reconstruction significantly improves the visualisation. For the LC, the low-resolution images cannot delineate substructures of the LC, which can be mistaken as high LC signal (Figure 6b vs 6d). Acquiring at a high-resolution clearly delineates this into two substructures. Figure 9b vs 9d shows that deep-learning reconstruction can improve SNR, however cannot resolve partial volume effects.

## Acknowledgements
Samy Abo Seada received funding from Parkinson NL. We would like to thank Professor Tom Ruigrok, from the department of neuroscience, Erasmus MC, for insightful discussions.